\documentclass[epj]{webofc}
\usepackage[utf8]{inputenc}
\usepackage[varg]{txfonts}   % Web of Conferences font
\usepackage{booktabs}
\usepackage{xcolor}
\definecolor{darkred}{rgb}{0.4,0.0,0.0}
\definecolor{darkgreen}{rgb}{0.0,0.4,0.0}
\definecolor{darkblue}{rgb}{0.0,0.0,0.4}
\usepackage[bookmarks,linktocpage,colorlinks,
    linkcolor = darkred,
    urlcolor  = darkblue,
    citecolor = darkgreen]{hyperref}
\usepackage{subfigure}
\wocname{EPJ Web of Conferences}
\woctitle{Lattice2017}
\newcommand{\bi}{\begin{itemize}}
\newcommand{\ei}{\end{itemize}}
\begin{document}
%%%%%%%%%%%%%%%%%%%%%%%%%%%%%%%%%%%%%%%%%%%%%%%%%%%%%%%%%%%%%%%%%%%%%%%%%%%%%
%
\selectlanguage{english}
%----------------------------------------------------------------------------
\title{%
Continuum extrapolation of critical point for finite temperature QCD with $N_{\rm f}=3$
}
%----------------------------------------------------------------------------
\author{%
\firstname{Xiao-Yong} \lastname{Jin}\inst{1} \and
\firstname{Yoshinobu} \lastname{Kuramashi}\inst{2,3} \and
\firstname{Yoshifumi}  \lastname{Nakamura}\inst{3}\and
\firstname{Shinji}  \lastname{Takeda}\inst{4}\fnsep\thanks{Speaker, \email{takeda@hep.s.kanazawa-u.ac.jp}}
\firstname{Akira}  \lastname{Ukawa}\inst{3}
% etc.
}
%----------------------------------------------------------------------------
\institute{%
Argonne Leadership Computing Facility, Argonne National Laboratory, Argonne, IL 60439, USA
\and
Center for Computational Sciences, University of Tsukuba, Tsukuba, Ibaraki 305-8577, Japan
\and
RIKEN Advanced Institute for Computational Science, Kobe, Hyogo 650-0047, Japan
\and
Graduate School of System Informatics, Department of Computational Sciences, Kobe University, Kobe, Hyogo 657-8501, Japan
\and
Institute of Physics, Kanazawa University, Kanazawa 920-1192, Japan
}
%----------------------------------------------------------------------------
\abstract{%
We study the critical point for finite temperature $N_{\rm f}=3$ QCD using several temporal lattice sizes up to 10.
In the study, the Iwasaki gauge action and non-perturbatively O(a) improved Wilson fermions are employed.
We estimate the critical temperature and the upper bound of the critical pseudo-scalar meson mass.
}
%----------------------------------------------------------------------------
\maketitle

%----------------------------------------------------------------------------

\section{Introduction}
\begin{figure}[t] % no figure before 1st section
  \centering
  \includegraphics[width=11cm,clip]{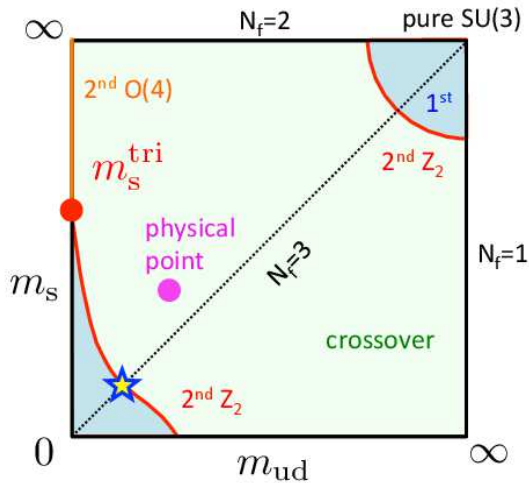}
  \caption{Columbia plot. Our aim is to fix the critical point (marked as star) on the 3-flavor symmetric line
  in the chiral region.}
  \label{fig:columbia}% Give a unique label
\end{figure}

It is an important issue to understand the nature of finite temperature QCD phase transition
with various parameters, for example, the quark masses and the number of flavors.
Such an information of the nature of the phase transition is summarized in the columbia plot (Fig.\ref{fig:columbia}).
The heavy region is well studied for example in Ref.\cite{Saito:2011fs}.
On the other hand, in the chiral region,
there are some qualitative predictions by the effective theory \cite{Pisarski:1983ms,Butti:2003nu,Calabrese:2004uk} with
the renormalization group analysis,
but the quantitative knowledge is still lacking.
Our ultimate goal is to quantitatively fix a shape of the critical line in the chiral region
by using the lattice QCD simulation.
See \cite{deForcrand:2006pv,Endrodi:2007gc} for previous studies on the critical line,
and \cite{Nakamura} for our current approach.
In this article, we focus on 3-flavor symmetric case and try to identify the critical point whose location is marked by a star in Fig.~\ref{fig:columbia}.

Table \ref{tab:} summarizes the value of the critical peseudo-scalar meson mass for 3-flavor QCD obtained
by the lattice simulations \cite{Aoki:1998gia,Karsch:2001nf,Liao:2001en,Karsch:2003va,deForcrand:2007rq,Varnhorst:2015lea,Bazavov:2017xul,Iwasaki:1996zt,Jin:2014hea,Takeda:2016vfj,Smith:2011pm,Jin:2017jjp}.
There are many works using staggered type fermions. 
Those results basically show that for larger $N_{\rm t}$ and more improvement, the critical mass tends to be small.
Recently the upper bound of the critical pion mass has been updated and it turns out to be very small
$m_{{\rm PS, critical}}\lesssim 50$ MeV \cite{Bazavov:2017xul},
and in some case it could be even zero \cite{Varnhorst:2015lea}.
On the other hand for Wilson type fermions, there is old study where the
critical mass was shown to be very heavy \cite{Iwasaki:1996zt}, but recently with the improvement of the action
we obtain a relatively smaller value.
In this article, we update our value of the critical mass by using larger temporal lattices \cite{Jin:2014hea,Takeda:2016vfj}.
More details of the contents in this article can be found in \cite{Jin:2017jjp}.

\begin{table}[t]
  \small
  \caption{Summary of the pseudo-scalar meson at the critical point for $N_{\rm f}=3$ QCD \cite{deForcrand:2017cgb}.}
  \label{tab:}% Give a unique label
  \centering
  \begin{tabular}{llll}\toprule
  Action  & $N_{\rm t}$ & $m_{{\rm PS,critical}}$ [MeV]& Ref.  \\\midrule
  staggered, standard & 4 & 290 & \cite{Karsch:2001nf}\\
  staggered, p4 & 4 & 67 & \cite{Karsch:2003va}\\
  staggered, standard & 6 & 150 & \cite{deForcrand:2007rq}\\
  staggered, HISQ & 6 & $\lesssim$ 50 & \cite{Bazavov:2017xul}\\
  staggered, stout & 4-6 & could be 0 & \cite{Varnhorst:2015lea}\\
  Wilson, standard & 4 & $\lesssim$ 670& \cite{Iwasaki:1996zt}\\
  Wilson, NP O($a$) improved & 4-8 & 300 & \cite{Jin:2014hea,Takeda:2016vfj}\\
  Wilson, NP O($a$) improved & 4-10 & $\lesssim 170$ & \cite{Jin:2017jjp}\\
  \end{tabular}
\end{table}

\section{Simulation setup}

We use the Iwasaki gauge action \cite{Iwasaki:2011np} and non-perturbatively O($a$) improved Wilson fermions \cite{Aoki:2005et}.
The temporal lattice $N_{\rm t}=4$, 6, 8 and 10 are used in our study.
To carry out the finite size scaling, the spatial lattice size is varied.
In order to determine the critical point we use the conventional kurtosis intersection analysis \cite{Karsch:2001nf}.
As an observable we use the naive chiral condensate.
We measure higher order of the mass-derivatives of the quark propagator by using the noise method.
They are used for the computation of the higher moments of the chiral condensate as well as
for the $\kappa$-reweighting, that is, the reweighting factor which is a ratio of quark determinant
is approximately computed as in \cite{Kuramashi:2016kpb}.
BQCD code \cite{Nakamura:2010qh} and RHMC algorithm \cite{Clark:2006fx} are used to generate the gauge configurations.

\section{Simulation results}

\begin{figure}[t] % no figure before 1st section
  \centering
\hspace{20mm}
  \includegraphics[width=12cm,clip]{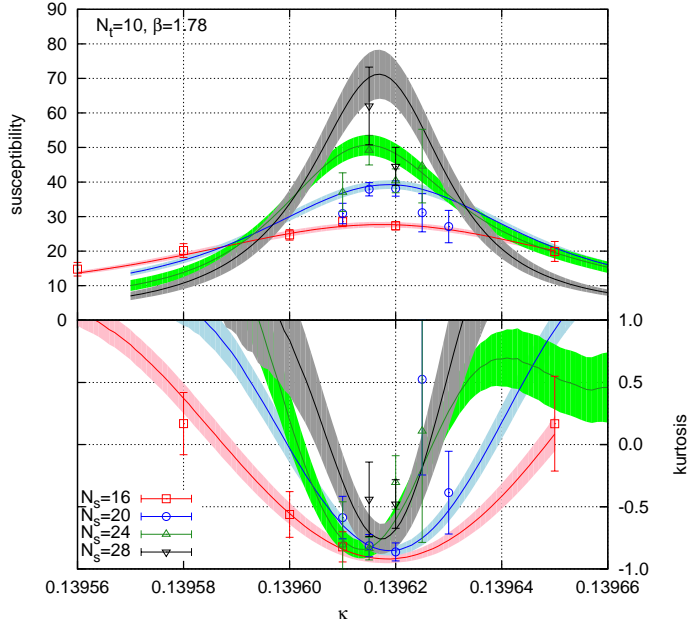}
  \caption{Susceptibility and kurtosis for the naive chiral condensate as a function of $\kappa$.
  Simulation parameters: $N_{\rm t}=10$, $\beta=1.78$, $N_{\rm s}=16$, $20$, $24$ and $28$.
  The band shows the multi-ensemble reweighting results.
  }
  \label{fig:1}% Give a unique label
\end{figure}

\begin{figure}[t] % no figure before 1st section
  \centering
\hspace{-5mm}
  \includegraphics[width=11.cm,clip]{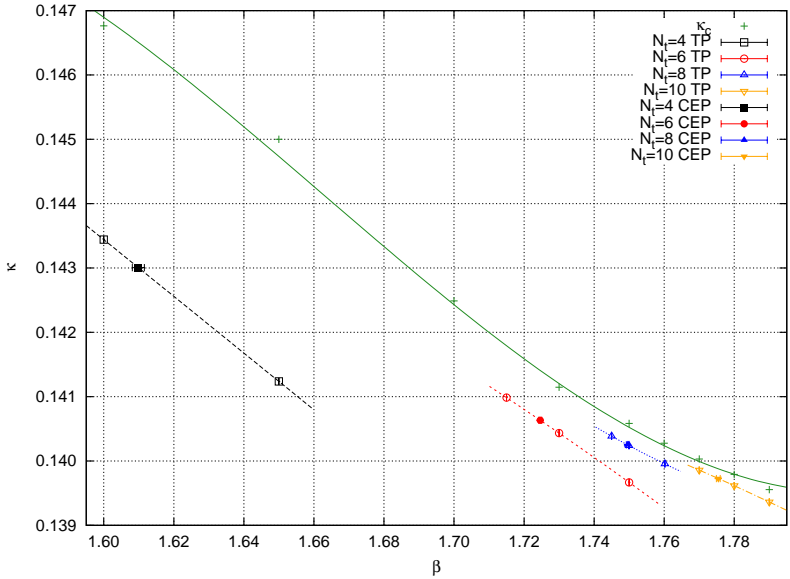}
  \caption{The
	transition points (TP) are represented by open symbols for $N_{\rm t}=4$, 6, 8 and 10.
	Filled symbols on the transition line show
	the location of the critical end point (CEP).
	The critical point is determined by the kurtosis intersection
	by using fitting formula in eq.(\ref{eqn:intersection}).
  The green curve represents the massless pion line at the zero-temperature.
  }
  \label{fig:2}% Give a unique label
\end{figure}

Typical results for the higher moments of
the chiral condensate (susceptibility and kurtosis)
are shown in Fig.~\ref{fig:1}
for $N_{\rm t}=10$ and $\beta=1.78$ and with some spatial volumes $N_{\rm s}=16-28$.
The kappa is used as a probe parameter.
In Fig.~\ref{fig:1}, $\kappa$-reweighting together with the multi-ensemble \cite{Ferrenberg:1988yz} results are also shown.
From the peak position of the susceptibility we determine the transition point.
Figure \ref{fig:2} summarizes the transition line for $N_{\rm t}=4$, $6$, $8$ and $10$
in the bare parameter space spanned by $\beta$ and $\kappa$.

\begin{figure}[t] % no figure before 1st section
  \centering
  \hspace{-7.2mm}
  \begin{tabular}{cc}
  \includegraphics[width=7cm,clip]{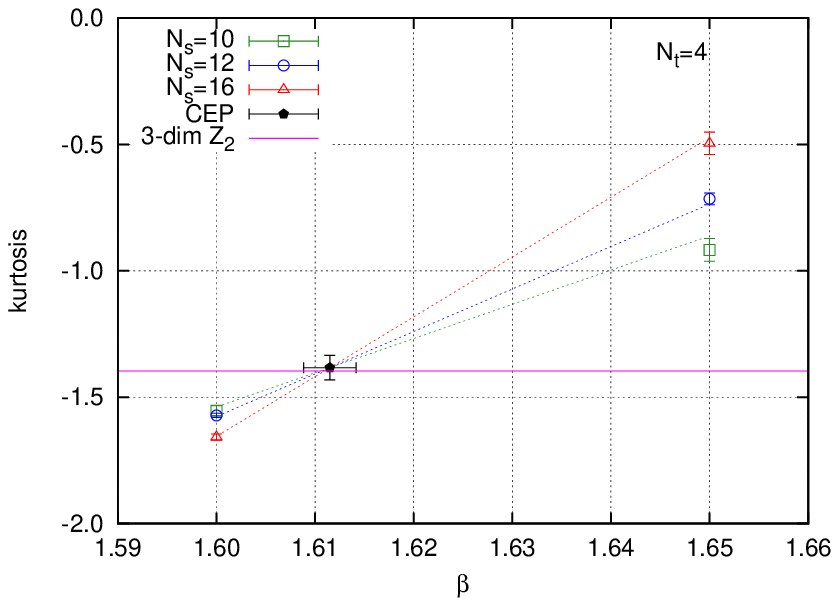}
&
  \includegraphics[width=7cm,clip]{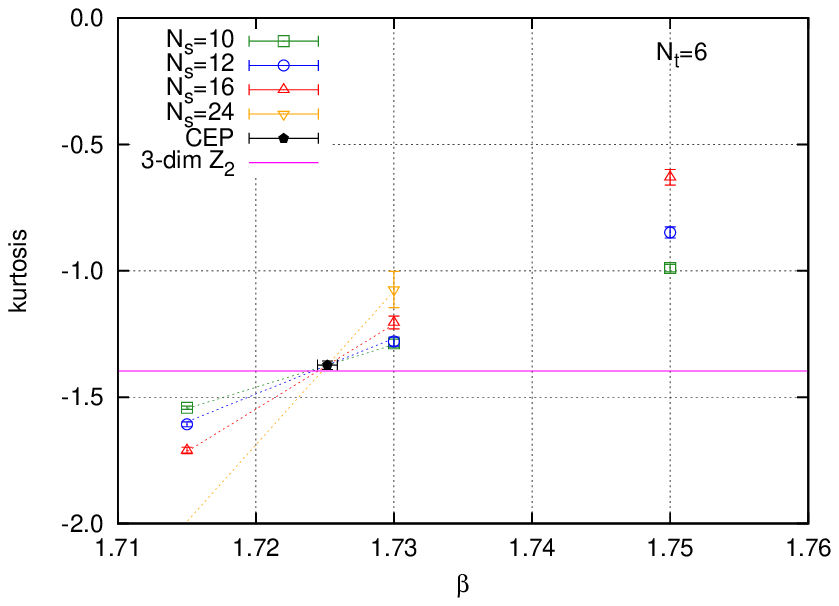}
\\
  \includegraphics[width=7cm,clip]{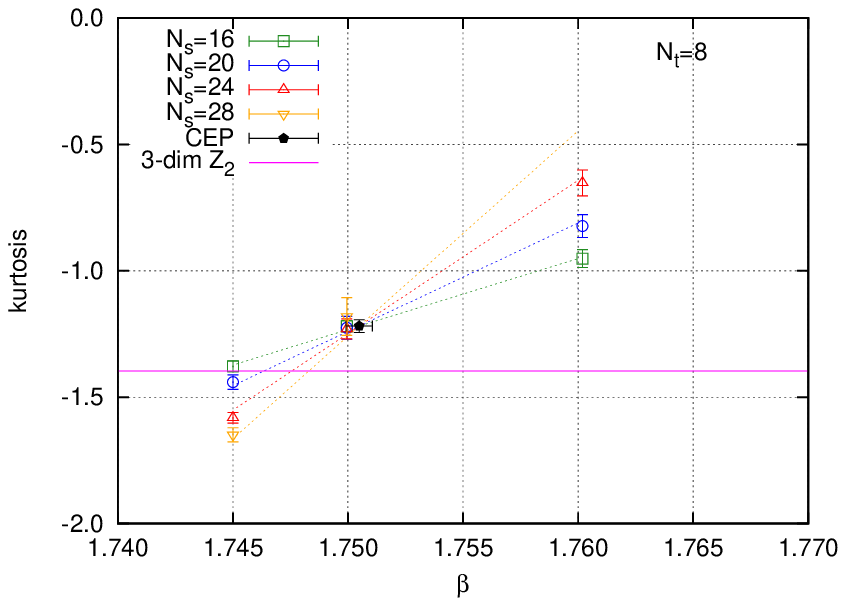}
&
  \includegraphics[width=7cm,clip]{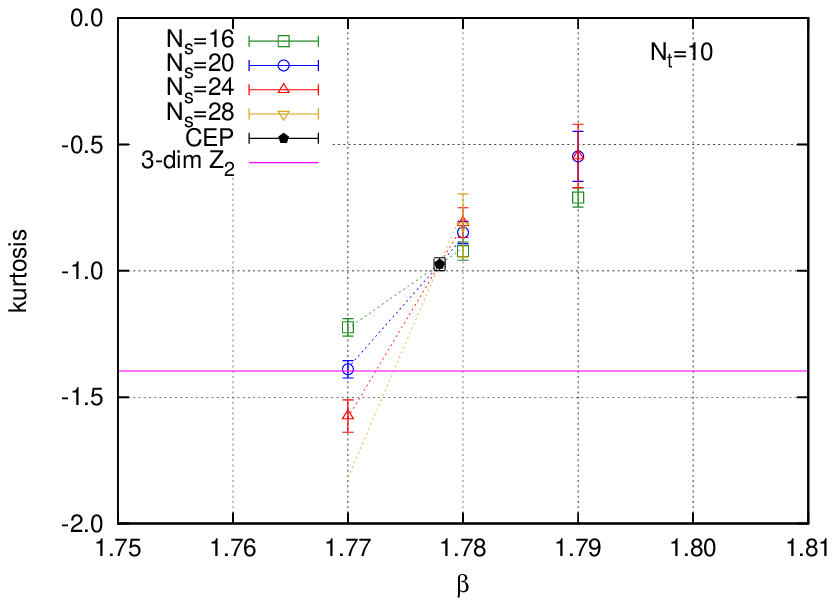}
\\
  \end{tabular}
  \caption{Kurtosis intersection for $N_{\rm t}=4$, $6$, $8$ and 10.
  The black point represents the intersection point.
  The horizontal magenta line shows the value of the kurtosis for three-dimensional
  Z$_2$ universality class.}
  \label{fig:3}% Give a unique label
\end{figure}

The minimum of kurtosis along the transition line projected on the $\beta$ axis is plotted in Fig.~\ref{fig:3}.
Although for smaller $N_{\rm t}$ ($=4$ and $6$) a consistency with the three-dimensional Z$_2$ universality class is clearly observed,
for higher $N_{\rm t}$ ($=8$ and $10$) the value of kurtosis of the critical point is larger than that of the Z$_2$.
One of possibility to explain this behavior is the finite size effects.
One conceives two sources of the finite size effects:
\begin{enumerate}
\renewcommand{\labelenumi}{(\Alph{enumi})}
\item Contribution of energy-like operator in the chiral condensate.
\item An effect of the leading irrelevant scaling field.
\end{enumerate}
Actually they suggest a similar modified form for the kurtosis intersection analysis,
\begin{equation}
K
=
\left[
K_{\rm E}
+
AN_{\rm s}^{1/\nu}(\beta-\beta_{\rm E})
\right]
(1+BN_{\rm s}^{-C}),
\label{eqn:intersection}
\end{equation}
where $C=C_{\rm A}=0.894$ for (A) and $C=C_{\rm B}=0.83...$ for (B) in the case of three-dimensional Z$_2$ universality class.
Since it is hard to distinguish between them in the fitting,
we fix $C=C_{\rm A}$ during the fitting procedure where $\beta_{\rm E}$, $A$ and $B$ are used as fitting parameters.
The fitting results are shown in Table \ref{tab:CEP}.
And then we obtain a reasonable result which is consistent with the Z$_2$ universality class, that is,
fitting assuming the value of $K_{\rm E}$ to be $-1.396$ and $\nu=0.63$ gives
reasonable $\chi^2/{\rm d.o.f.}\sim O(1)$ for $N_{\rm t}=8$ and 10.
The resulting new critical points are plotted in Fig.~\ref{fig:2}.
%In the following, we use the new critical point determined by the modified fitting formula.

%%%%%%%%%
\begin{table}[t]
  \footnotesize
\caption{
Fit results for kurtosis intersection with fitting form in eq.(\ref{eqn:intersection}).
%See text for the definition of Fit-1, 2 and 3.
%Fit-1: no correction term ($B=y_{\rm t}-y_{\rm h}=0$) and all other parameters are used as fit parameter.
%Fit-2: no correction term and assuming the 3D Z$_2$ universality class for $K_{\rm E}$ and $\nu$.
%Fit-3: including correction term and assuming the 3D Z$_2$ universality class for $K_{\rm E}$, $\nu$ and $y_{\rm t}-y_{\rm h}$.
A value without error bar means that the corresponding fit parameter is fixed to the given value during the fit.
%For the 3D Z$_2$ universality class, the expected values of the parameter are
%$K_{\rm E}=-1.396$, $\nu=0.630$ and $C=0.894$ respectively.
%Using the value of $\beta_{\rm E}$ as an input,
%$\kappa_{\rm E}$ is obtained from an interpolation formula of the transition line in Fig.~\ref{fig:phase_diagram}.
}
\label{tab:CEP}
\hspace{-2mm}
  \centering
\begin{tabular}{rrllllllrr}\toprule
$N_{\rm t}$
&
Fit
&
$\beta_{\rm E}$
&
$\kappa_{\rm E}$
&
$K_{\rm E}$
&
$\nu$
&
$A$
&
$B$
&
$C$
&
$\chi^2/{\rm d.o.f.}$
\\
\hline
4&
$1$&$1.6115 ( 26 )$&$0.1429337 ( 13 )$&$-1.383 ( 48 )$&$0.84 ( 13 )$&$0.88 ( 42 )$&$\times$&$\times$&$1.75$\\
&
$2$&$1.61065 ( 61 )$&$0.1429713 ( 13 )$&$-1.396$&$0.63$&$0.313 ( 12 )$&$\times$&$\times$&$3.05$\\
&
$3$&$1.6099 ( 17 )$&$0.1430048 ( 13 )$&$-1.396$&$0.63$&$0.311 ( 14 )$&$0.10 ( 21 )$&$0.894$&$3.77$\\
\hline
6&
$1$&$1.72518 ( 71 )$&$0.1406129 ( 14 )$&$-1.373 ( 17 )$&$0.683 ( 54 )$&$0.58 ( 17 )$&$\times$&$\times$&$0.68$\\
&
$2$&$1.72431 ( 24 )$&$0.1406451 ( 14 )$&$-1.396$&$0.63$&$0.418 ( 11 )$&$\times$&$\times$&$0.70$\\
&
$3$&$1.72462 ( 40 )$&$0.1406334 ( 14 )$&$-1.396$&$0.63$&$0.422 ( 12 )$&$-0.052 ( 52 )$&$0.894$&$0.70$\\
\hline
8&
$1$&$1.75049 ( 57 )$&$0.1402234 ( 11 )$&$-1.219 ( 25 )$&$0.527 ( 55 )$&$0.146 ( 88 )$&$\times$&$\times$&$0.73$\\
&
$2$&$1.74721 ( 42 )$&$0.14031921 ( 76 )$&$-1.396$&$0.63$&$0.404 ( 36 )$&$\times$&$\times$&$5.99$\\
&
$3$&$1.74953 ( 33 )$&$0.1402512 ( 10 )$&$-1.396$&$0.63$&$0.414 ( 13 )$&$-1.33 ( 15 )$&$0.894$&$0.73$\\
\hline
10&
$1$&$1.77796 ( 48 )$&$0.1396661 ( 17 )$&$-0.974 ( 25 )$&$0.466 ( 45 )$&$0.084 ( 52 )$&$\times$&$\times$&$0.22$\\
&
$2$&$1.7694 ( 16 )$&$0.1398724 ( 22 )$&$-1.396$&$0.63$&$0.421 ( 95 )$&$\times$&$\times$&$10.03$\\
&
$3$&$1.77545 ( 53 )$&$0.1397274 ( 17 )$&$-1.396$&$0.63$&$0.559 ( 29 )$&$-2.97 ( 25 )$&$0.894$&$0.43$\\
\end{tabular}
\end{table}
%%%%%%%%%

\section{Continuum extrapolation of the critical point}
\begin{figure}[t] % no figure before 1st section
  \centering
\hspace{-5mm}
  \includegraphics[width=10cm,clip]{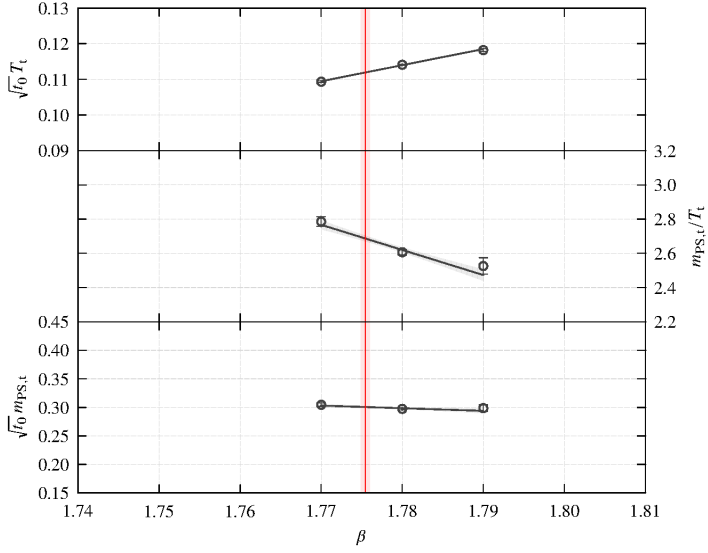}
  \caption{The transformation of the critical point from the bare parameter space to physical one.
  The red vertical line shows the location of the bare critical point.
  The black line is the interpolation of the dimensionless physical quantity along the transition line.
  The intersection of the red and black line gives the estimation of $\sqrt{t_0}T$, $m_{\rm PS}/T$ and $\sqrt{t_0}m_{\rm PS}$ at the critical point.}
  \label{fig:4}% Give a unique label
\end{figure}

Next step is to express the bare critical point in terms of hadronic physical quantity.
For that purpose we carry out the zero temperature simulation and compute the Wilson flow scale $\sqrt{t_0}$ \cite{Luscher:2010iy} and the
pseudo-scalar meson mass $m_{\rm PS}$.
The transformation of the critical point from the bare parameter space to the physical one for $N_{\rm t}=10$
is shown in Fig.~\ref{fig:4}.

\begin{figure}[b] % no figure before 1st section
  \centering
  \begin{tabular}{cc}
  \includegraphics[width=6.5cm,clip]{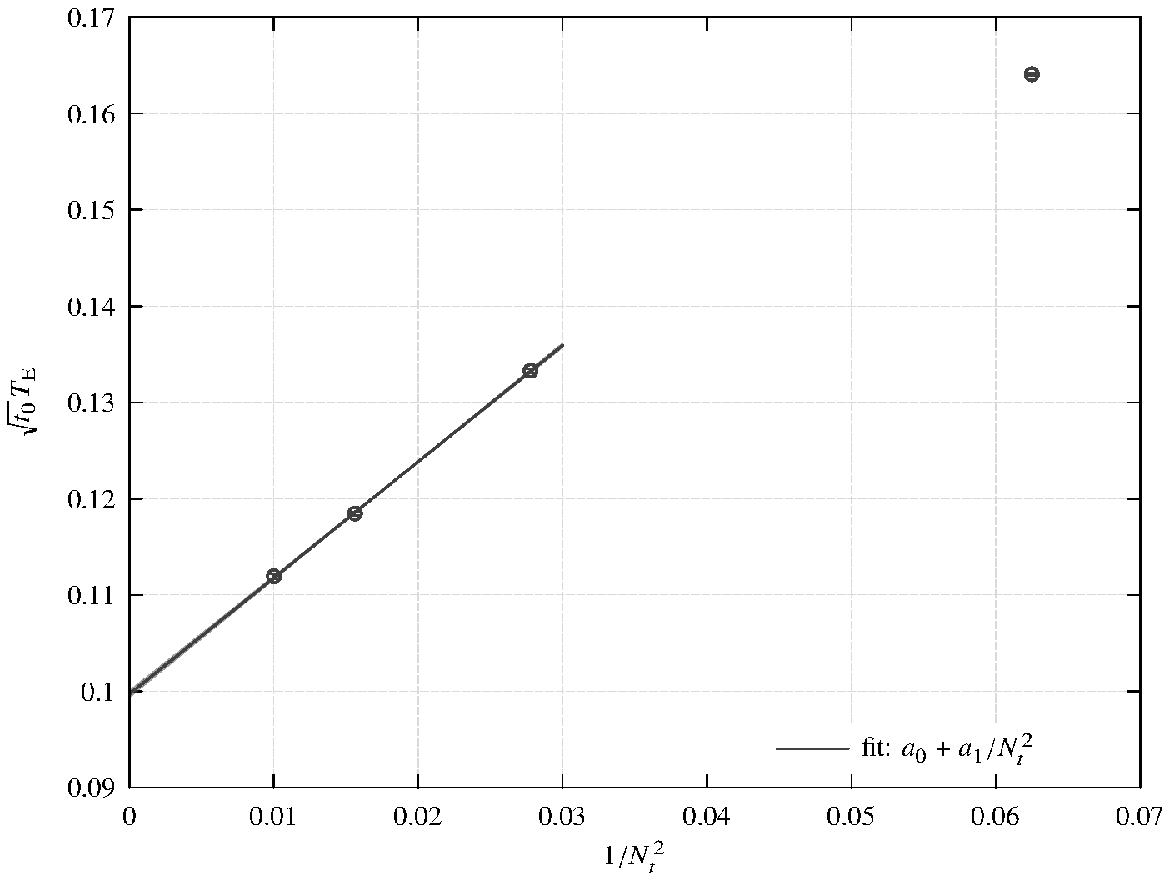}
&
  \includegraphics[width=6.5cm,clip]{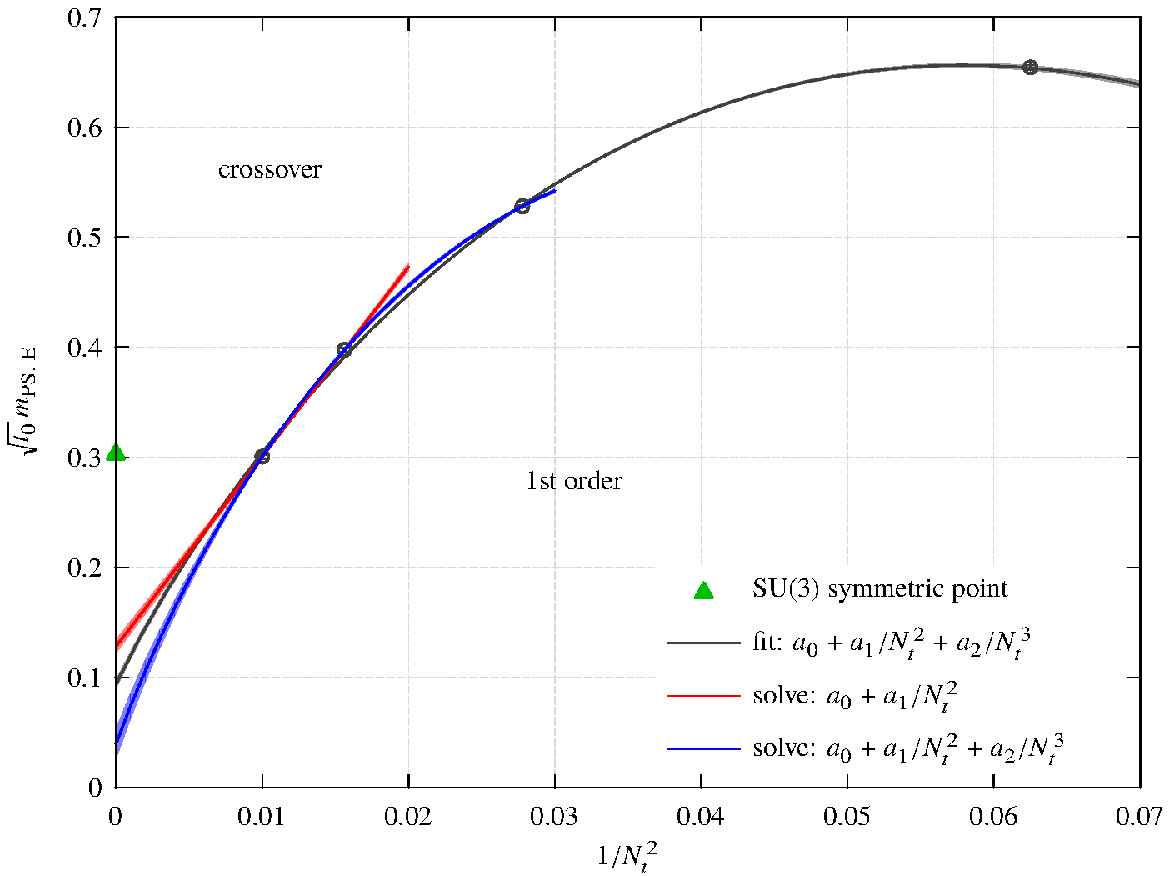}
	\\
  \includegraphics[width=6.5cm,clip]{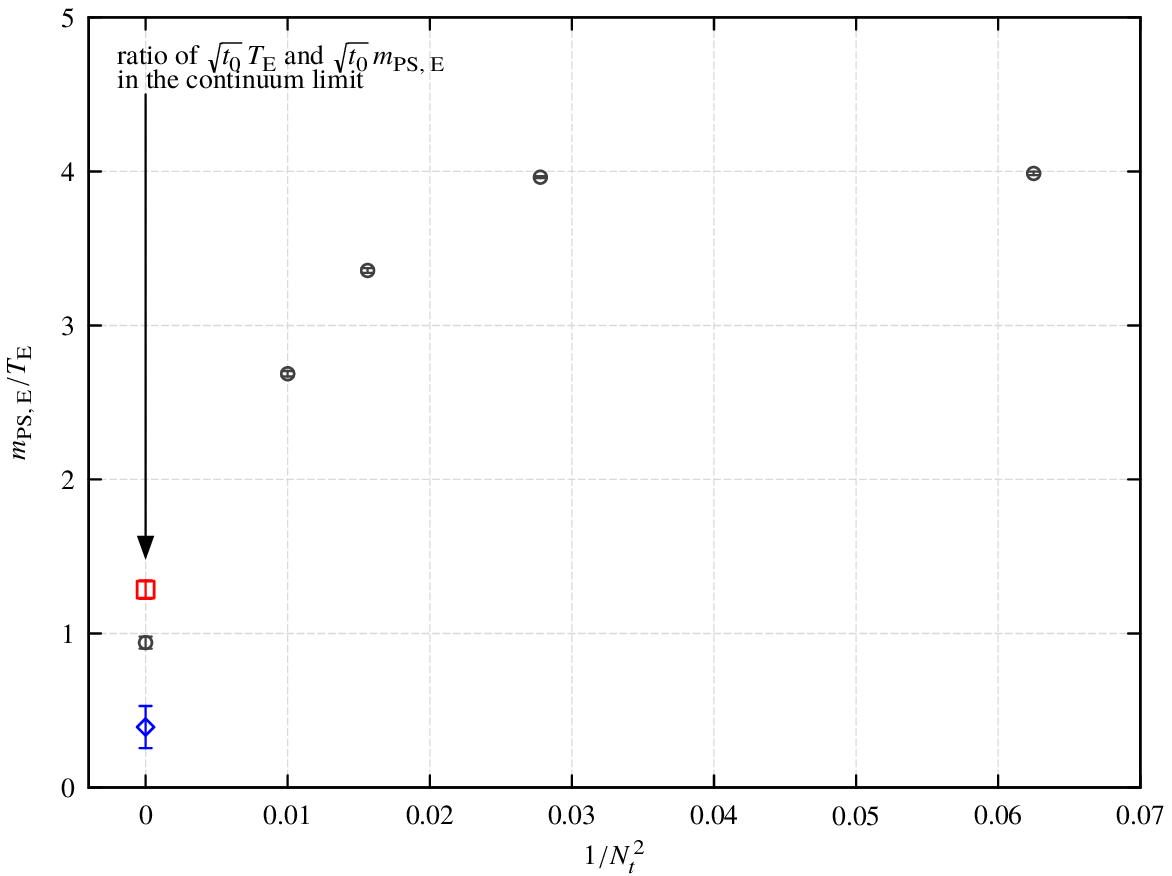}
&
  \includegraphics[width=6.5cm,clip]{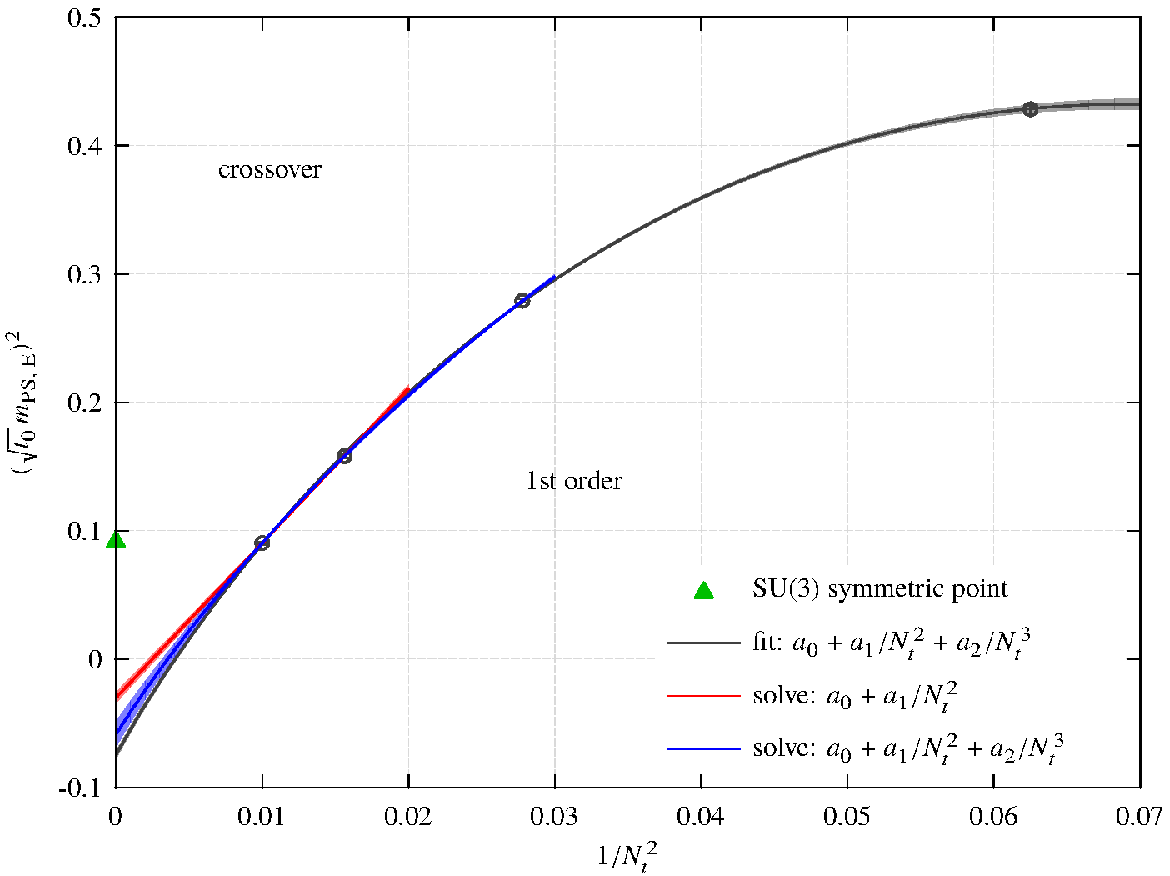}
	\\
	\end{tabular}
  \caption{Continuum extrapolation of the critical temperature and the critical pseudo-scalar meson mass.}
  \label{fig:5}% Give a unique label
\end{figure}

The continuum extrapolation of the critical temperature $T_{\rm E}$ is shown in Fig.~\ref{fig:5} (upper-left panel).
The data points at $N_{\rm t}=6$, 8 and 10 are in a good scaling region and we obtain the continuum value (in physical units \cite{Borsanyi:2012zs})
\begin{equation}
T_{\rm E}=134(3)\mbox{MeV}.
\end{equation}

On the other hand, for critical pion mass in Fig.~\ref{fig:5} (upper-right and lower-left), it has a large
scaling violation and the extrapolation has ambiguity, namely the extrapolation depends on fitting range and fitting form.
We also check the pion mass square as in Fig.\ref{fig:5} (lower-right).
In this case, we somehow obtain a negative value.
This shows that our data points may be not in the scaling region
and it is hard to quote a single value for an extrapolation.
Therefore, we conservatively estimate an upper bound of the critical pion mass.
We take the maximum value among all fits that we did as the upper bound (in physical units),
\begin{equation}
m_{\rm PS,E} \lesssim 170\mbox{MeV}.
\end{equation}

\begin{figure}[t] % no figure before 1st section
  \centering
\hspace{-5mm}
  \includegraphics[width=7.5cm,clip]{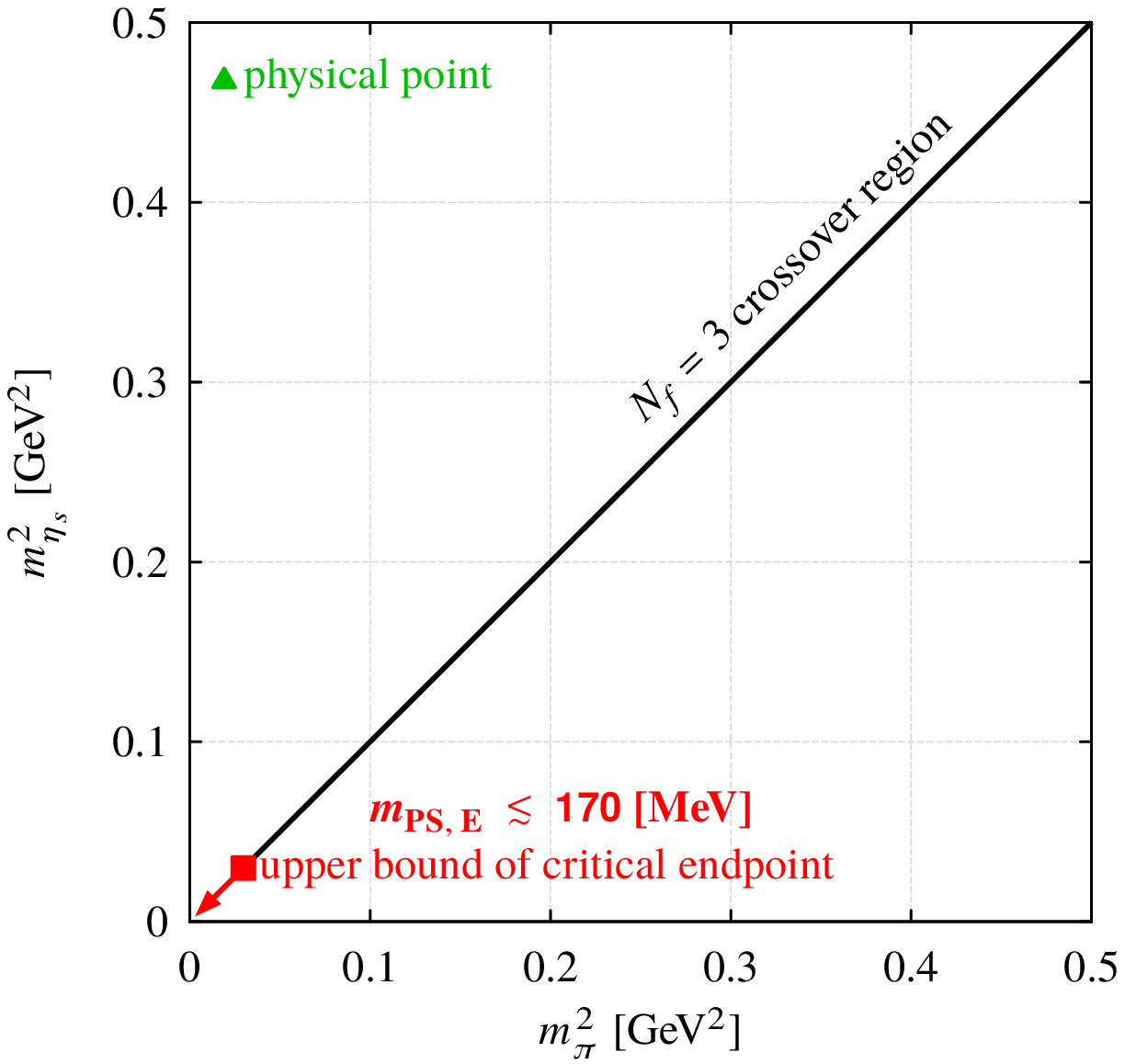}
  \caption{Summary plot (Columbia-like plot) of our result.}
  \label{fig:6}% Give a unique label
\end{figure}

\section{Summary and outlook}
We carry out the finite temperature $N_{\rm f}=3$ QCD simulation by using non-perturvatively O($a$) improved Wilson fermions
and study the critical end point by using the kurtosis intersection.
By carrying out large scale simulations up to $N_{\rm t}=10$, we attempt to take
the continuum limit of the critical end point.
Although we stably obtain the critical temperature $T_{\rm E}=134(3)$ MeV,
the critical pseudo-scalar meson mass has large ambiguity caused in the continuum extrapolation.
Therefore we conservatively estimate the upper bound $m_{\rm PS,E}\lesssim 170$ MeV.
A summary plot of our result is shown in Fig.~\ref{fig:6} where
both axes are represented by meson masses $m_\pi^2$ and $m_{\eta_{\rm s}}^2$ which may be proportional to the quark masses
$m_{\rm ud}$ and $m_{\rm s}$ respectively.
Our upper bound is relatively larger than that of HISQ \cite{Bazavov:2017xul}, $m_{\rm PS,E}\lesssim50$ MeV.
Note that, however, our upper bound is estimated from the fact that there is critical point,
while in the HISQ study the bound comes from the absence of the critical point.

In future, it is very important to fix the value of the critical end point or push down the upper bound.
For Wilson type fermions, one has to handle the scaling violation.
In order to take the continuum limit more smoothly and reliably one needs to run further large scale simulations say at $N_{\rm t}=12$
or implement some improvement on the lattice action or lattice setup.

%\clearpage

\section*{Acknowledgements}
This research used computational resources of 
HA-PACS and COMA provided by Interdisciplinary Computational Science Program in Center for Computational Sciences at University of Tsukuba,
System E at Kyoto University through the HPCI System Research project (Project ID:hp150141),
PRIMERGY CX400 tatara at Kyushu University
and
HOKUSAI GreatWave (project ID:G16016) at RIKEN.
This work is supported by JSPS KAKENHI Grant Numbers 26800130,
FOCUS Establishing Supercomputing Center of Excellence.
This research used resources of the Argonne Leadership Computing Facility, which is a DOE Office of Science User Facility supported under Contract DE-AC02-06CH11357.

%\clearpage
%\bibliography{lattice2017}

%%%%%%%%%%%%%%%%%%%%%%%%%%%%%%%%%%%%%%%%%%%%%%%%%%%%%%%%%%%%%%%%%%%%%%%%%%%%%
\end{document}